\documentclass[preprint,12pt,authoryear]{elsarticle}

\usepackage{amssymb}
\usepackage{lipsum}
\usepackage{csquotes}
\usepackage[colorlinks = true, linkcolor = blue, urlcolor  = blue, citecolor = blue]{hyperref}

\usepackage{amsmath}
\newcommand{\rick}{\texttt{RICK}}
\newcommand{\CU}{\texttt{CU}}
\newcommand{\CUs}{\texttt{CUs}}

\journal{Astronomy $\&$ Computing}

\begin{document}

\begin{frontmatter}



\title{Accelerating radio astronomy imaging with RICK: a step towards SKA-Mid and SKA-Low}


\author[OABR,ICSC]{G. Lacopo}
            
\author[UNIBO,IRA,ICSC]{E. De Rubeis}

\author[IRA,ICSC]{C. Gheller}

\author[OATS,ICSC]{G. Taffoni}

\author[OATS,ICSC]{L. Tornatore}

\affiliation[OABR]{organization={INAF-Osservatorio astronomico di Brera},%Department and Organization
            addressline={Via E. Bianchi, 46}, 
            city={Merate},
            postcode={23807}, 
            state={LC},
            country={Italy}}
\affiliation[ICSC]{organization={ ICSC -- Centro Nazionale di Ricerca in High Performance Computing, Big Data e Quantum Computing},%Department and Organization
            addressline={via Magnanelli 2}, 
            city={Bologna},
            postcode={40033}, 
            state={TS},
            country={Italy}}
\affiliation[OATS]{organization={INAF-Osservatorio Astronomico di Trieste},%Department and Organization
            addressline={via Tiepolo 11}, 
            city={Trieste},
            postcode={34143}, 
            state={TS},
            country={Italy}}
\affiliation[UNIBO]{organization={Dipartimento di Fisica e Astronomia, Università di Bologna},%Department and Organization
            addressline={via Gobetti 93/2}, 
            city={Bologna},
            postcode={I-40129}, 
            state={BO},
            country={Italy}}
\affiliation[IRA]{organization={Istituto di Radioastronomia, INAF},%Department and Organization
            addressline={via Gobetti 101}, 
            city={Bologna},
            postcode={40129}, 
            state={BO},
            country={Italy}}

\begin{abstract}
The data volumes generated by modern radio interferometers, such as the SKA precursors, present significant computational challenges for imaging pipelines. Addressing the need for high-performance, portable, and scalable software, we present \rick~2.0 (Radio Imaging Code Kernels). This work introduces a novel implementation that leverages the HeFFTe library for distributed Fast Fourier Transforms, ensuring portability across diverse HPC architectures, including multi-core CPUs and accelerators. We validate \rick's correctness and performance against real observational data from both MeerKAT and LOFAR. Our results demonstrate that the HeFFTe-based implementation offers substantial performance advantages, particularly when running on GPUs, and scales effectively with large pixel resolutions and a high number of frequency planes. This new architecture overcomes the critical scaling limitations identified in previous work (Paper II, Paper III), where communication overheads consumed up to $96\%$ of the runtime due to the necessity of communicating the entire grid. This new \rick~version drastically reduces this communication impact, representing a scalable and efficient imaging solution ready for the SKA era.
\end{abstract}




\end{frontmatter}

%\tableofcontents



\section{Introduction}
\label{sec:Introduction}
Interferometric imaging in radio astronomy pipelines is one of the most computationally challenging steps, with algorithmic developments that must catch up with the surveys that now provide up to petabytes of data per year, such as International LOw Frequency ARray (LOFAR) Telescope \citep[ILT,][]{van2013lofar}, Atacama Large Millimeter Array \citep[ALMA,][]{wootten2009atacama}, MeerKAT \citep{jonas2016meerkat}, Murchison Widefield Array \citep[MWA,][]{tingay2013murchison}, Australian SKA Pathfinder Telescope \citep[ASKAP,][]{johnston2008science}, and eventually Square Kilometre Array \citep[SKA,][]{schaubert2003square}, which is supposed to produce terabytes of data per second, meaning hundreds of petabytes per year. MeerKAT, ASKAP and MWA are precursors, meaning that they will be extended and included in what will become the SKA. The others are pathfinders, meaning that they are engaged in SKA-related technology and science studies.

Handling such a large amount of data is an outstanding challenge in computer science, because they are not supposed to fit in filesystems and ``real-time computing" is unavoidable, i.e. at one point the filesystem memory runs out and data must be deleted in order to leave space for the new ones. This means that many relevant scientific data will be lost if the imaging pipeline is not quick enough. In addition, radio astronomy data is packed into Measurement Sets, which are read thanks to the Casacore software \citep{van2015casacore}. The Measurement Set format, although being historically very useful, does not scale particularly well and often the science process requires non-optimal access, giving rise to additional I/O load \citep{dodson2024optimising}. Today, many codes are available to perform all imaging steps, like WSClean \citep{offringa-wsclean-2014}, CASA \citep{bean2022casa} and DDFacet \citep{monnier2022multi}. 

However, handling extraordinarily large datasets often presents limitations concerning support for full distributed parallelism via the Message Passing Interface (MPI). Consequently, single-node parallelization is frequently constrained to multi-threaded methodologies, leveraging frameworks such as OpenMP or Pthreads. While certain code segments have been successfully accelerated by porting to Graphics Processing Units (GPUs), a critical limitation remains the lack of support for multi-GPU architectures. Furthermore, the memory capacity of a single GPU is typically insufficient to accommodate these substantial data volumes. Given that these datasets frequently exceed the collective memory of a single compute node, they necessitate decomposition into sequential temporal or frequency-domain chunks for iterative processing, invariably leading to pronounced I/O bottlenecks.

All these necessities led us to develop the RICK (Radio Imaging Code Kernels) code, presented in \citet{gheller2023}, hereafter Paper I, \citet{de2025accelerating}, hereafter Paper II, \citet{lacopo2025green}, hereafter Paper III, and \citet{de2025hpc}. RICK has been thought for massively parallel machines, with GPU offloading through CUDA \citep{farber2011cuda}, HIP\footnote{\url{https://rocm.docs.amd.com/projects/HIP/en/latest/what_is_hip.html}} and OpenMP for GPGPU \citep{lee2009openmp,deakin2023programming}. The MPI+GPU version supported NCCL\footnote{\url{https://developer.nvidia.com/nccl}} for GPU-GPU communication, and cuFFTMP\footnote{\url{https://docs.nvidia.com/cuda/cufftmp/index.html}} for FFT calculation on NVIDIA GPUs. In Paper II we demonstrated that running RICK in full (NVIDIA) GPU configuration was two orders of magnitude faster than the MPI+OpenMP CPU implementation. Tests for portability on AMD GPUs with time-to-solution and the corresponding energy-to-solution between CPU and GPU configurations have been carried out and discussed in Paper III. Nevertheless, all papers stress that in RICK MPI communication introduces a strong overhead, prevents the code for scaling and turns out to take up to $\sim 95\%$ of runtime. The effect gets dramatic when a purely distributed approach is used and the image pixel resolution gets large.

The communication bottleneck, which has been extensively discussed in \citet{phd_thesis_lacopo}, led us to develop the RICK 2.0 version, in which the communication impact is minimized. The preceding \rick~version was very good when a few number of MPI tasks were used, but the communication impact, which was due to an allreduce operation, was becoming dramatic when the code was run on many computing nodes and large images needed to be produced, especially in pure MPI runs, i.e. without OpenMP threads and/or GPUs. The development of a code version which handles such large problems is the main reason of the present paper, which describes the new implementation of RICK. The data structure has been entirely redefined, with data being redistributed as a first step, such that data locality through all the successive pipeline steps is guaranteed. To handle such data structure, a modern FFT library has been implemented, and I/O in reading data and writing output have been redesigned with MPI-I/O. 

From the scientific point of view, RICK validation against MeerKAT and LOFAR observations has been carried out and results are being presented in this paper. The radio imaging problem is described in Section \ref{sec:w-stacking}, the RICK 2.0 implementation is the subject of Section \ref{sec:Implementation}, and methodology and results are discussed in Section \ref{sec:methodology}. Conclusions are drawn in Section \ref{sec:conclusions}.

\section{The $w$-stacking gridder}
\label{sec:w-stacking}

\begin{figure}
\includegraphics[width=\textwidth]{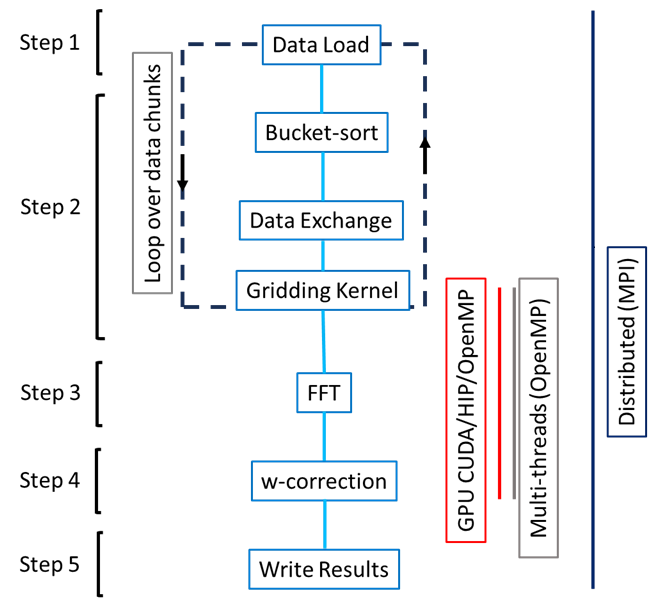}
\caption{Schematic code architecture and workflow of \rick, based on the one in~\citet{gheller2023} with the new steps that we ported on GPUs (reduce and FFT). Different kind of HPC enabling are highlighted with different colours.}
\label{fig:workflow}
\end{figure}

An interferometer measures complex visibilities $V$ related to the sky brightness distribution $I$ as:
\begin{equation}
\begin{split}
V(u,v,w) = &\int\int \frac{I(l,m)}{\sqrt{1-l^2-m^2}} \times \\ 
         &e^{-2\pi i \left(ul + vm + w(\sqrt{1-l^2-m^2}-1)\right)} dl dm,
\label{eq:visI}
\end{split}
\end{equation}
$l$ and $m$ are the sky coordinates, and $u$, $v$, and $w$ are the baseline coordinates in units of $\lambda$, with $w$ chosen to be in the direction of the source. The $u$,$v$ plane coverage is the distribution of baselines in $\lambda$ units as seen from the source at infinity, where each projected baseline corresponds to a point with coordinates $u$,$v$ in the Fourier space.
For small fields of view, the term $\sqrt{1-l^2-m^2}$ is close to unity, and Eq.~\ref{eq:visI} reduces to an ordinary two-dimensional Fourier transform, which can be efficiently computed using a Fast Fourier Transform (FFT) approach.
When large Fields of View (FoV) are observed, visibility data from non-coplanar interferometric radio telescopes cannot be accurately imaged with a two-dimensional Fourier transform, and the imaging algorithm needs to account for the $w$-term. One approach to account for the $w$ term is the $w$-stacking method $w$ direction, and visibilities are mapped to the closest w-plane:
\begin{equation}
\begin{split}
\frac{I(l,m)\left(w_{\max} - w_{\min}\right)}{\sqrt{1-l^2-m^2}} = &\int\limits_{w_{\min}}^{w_{\max}} e^{2\pi i w(\sqrt{1-l^2-m^2}-1)} \times 
\\
&\iint V(u,v,w)  e^{2\pi i \left(ul + vm\right)} du dv dw.
\label{eq:wstacking}
\end{split}
\end{equation}
After gridding, each w-plane is Fourier transformed separately, and a correction is applied as:
$w$-projection~\citep{2008ISTSP...2..647C} due to its algorithmic characterization, especially when gridding is the dominant cost~\citep[see Tab.~1 in][]{offringa-wsclean-2014}, resulting in slightly lower image errors.
The $w$-stacking algorithm has been implemented in Paper I within the $w$-stacking gridder. Two main data structures characterize the algorithm: an unstructured dataset storing the $(u,v,w)$ coordinates of the antennas array baselines at each measurement time, and a Cartesian computational mesh of size $N_u \times N_v \times N_w$. The convolved visibilities and their FFT transformed counterpart are calculated on this mesh.
The code consists of five main algorithmic components (shown in Fig.~\ref{fig:workflow}), each supporting different types of HPC implementations. The components are:
\begin{itemize}
    \item Parallel data reading and distribution to MPI tasks.
    \item Gridding of visibilities, which involves creating an array for each slab, concatenating data with $u-v$ coordinates, convolving with a kernel (in this case, a Kaiser-Bessel function $G$ \citep{jackson_91}), and exchanging boundary data among slabs, here are the mathematical details:
    \begin{equation}
    \tilde V(u_i,v_j,w_k) = \sum_{m \in {\rm measures}} V_m G((u_m,v_m,w_m),(u_i,v_j,w_k)),
    \label{eq:convolution}
    \end{equation}
    where $m$ is the $m$-th measurement, $(u_m,v_m,w_m)$ are its coordinates, $(u_i,v_j,w_k)$ is a computational grid point, $V_m$ is the measured visibility and $\tilde V$ is the visibility convolved on the mesh.

    \item FFT of the gridded data.
    \item Phase shift and reduction of w-planes.
    \item Parallel writing of final images.
\end{itemize}

\section{Implementation}
\label{sec:Implementation}

The code has undergone a comprehensive refactoring compared to its previous version, which was extensively discussed in Papers I, II, and III. In summary, in the original version with P processes and T hours of observation, the first $\frac{T}{P}$ hours would be assigned to process $P_0$, the next segment to process $P_1$, and so forth. The two-dimensional mesh (the grid) is divided into regular rectangular slabs, with each process handling a distinct slab in the $u-v$ plane. The visibilities and grid constitute the primary memory requirements of \rick, and by distributing these among all MPI processes, the code is theoretically capable of handling problems of arbitrary size. This section will outline the primary modifications to the code previously presented in Papers I, II, and III, aimed at reducing the communication.

The solutions that we addressed to enhance the code efficiency and eliminate the MPI Reduce are:
\begin{itemize}
    \item a novel parallel I/O approach based on the MPI library is employed for reading data from the filesystem and writing the final image.
    \item A two-dimensional grid with a non-uniform distribution is initialized, reflecting the non-uniform distribution of visibilities in the $u-v$ plane, where central regions exhibit a significantly higher density of points.
    \item Visibilities are assigned positions on the grid according to their $v$ coordinates and subsequently bucket-sorted into respective grid slabs.
    \item MPI Sendrecv operations are utilized to communicate relevant data to each MPI process for its assigned slab.
    \item With data localized to each MPI process, the gridding step is executed, benefiting from load balancing due to varying visibility densities across slabs.
    \item A new FFT implementation is required to accommodate non-uniform domain decompositions.
    \item The $w$-correction term is computed using the same algorithm, albeit with a non-uniform grid.
\end{itemize}

\subsection{Parallel I/O}
\label{parallel-io-theory}
MPI-I/O has been implemented in \rick, which used \textit{fopen}, \textit{fread} and \textit{fseek} with an offset associated with each MPI task, when data were read from binary measurement sets, as discussed in \citet{2024SPIE13101E..1HG}. MPI-I/O supports both parallel and non-parallel filesystems and guarantees a better scalability than the previous version. A revised version of the function:
\begin{verbatim}
MPI_File_read_at(MPI_File fh, MPI_Offset offset, void *buf,
		     int count, MPI_Datatype datatype, MPI_Status *status);
\end{verbatim}
was needed, in order to carry out the I/O buffers in loop with maximum dimension equal to the integer max, which is the maximum value for 32-bit signed integers in C standard; otherwise, count to be the maximum integer limits the applicability of the code when many gigabytes of visibilities are read and few MPI tasks are used. Thanks to MPI datatypes, this caveat has been overcome. 

MPI-I/O has also been implemented in writing output, because the preceding \rick~version used a different approach based on MPI shared windows to write files containing intermediate results, often needed for debugging reasons. Eventually, MPI-I/O has been included in the final image writing, whereas previously each MPI rank wrote its own sector with \textit{fwrite}, but it happened sequentially with semaphores. The improvement in using MPI-I/O guarantees portability, that was not full in the \rick~version presented in Paper I, II and III, which used to need a parallel filesystem.

\subsection{Domain Decomposition}
\label{dd-theory}
To avoid the communication of the entire grid, that was needed in the original \rick~version to assign each MPI rank its sector in $u-v$ plane, a domain decomposition rearrangement must now be carried out. The computation was evenly distributed among the processes, and halved by doubling the computing resources, nevertheless the grid to be communicated was always the same and could not led to expected strong and weak scaling (see Paper II). 

\begin{figure}[h]
    \centering
    \includegraphics[width=0.7\textwidth]{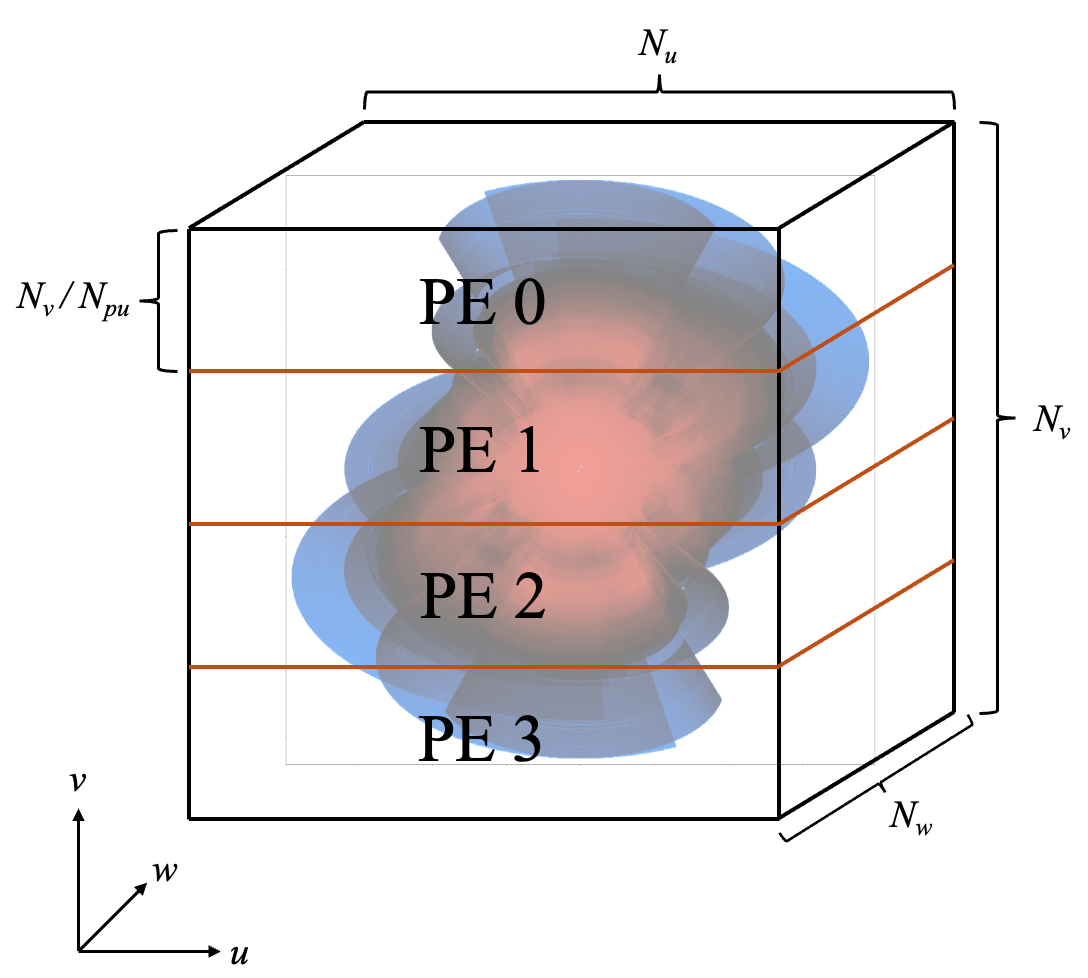}
    \caption{Domain decomposition in rectangular slabs along the $v$ axis, with four MPI processes. Visibilities are originally distributed in time-order, whereas the grid is initialized in space-order, requiring an all-to-all communication once the gridding step is carried out.}
    \label{fig:original_dd}
\end{figure}

In version 1.0, each MPI rank read data from the measurement set in time-order. Then, the gridding step was carried out from each process on its time-ordered data, and after that, the grid communication was performed. This strategy led to a double memory occupancy, since each process had to allocate the grid twice, once for gridding and once for reduce, eventually focusing only on the final grid, after the communication completion. The grid domain decomposition was one-dimensional (along the $v$ axis), and evenly distributed among the processes, i.e. no unbalance was present neither in gridding nor in Fast Fourier Transform steps. Figure \ref{fig:original_dd} is the three-dimensional generalization of the domain decomposition, where rectangles become parallelepipeds when the direction $w$ is taken into account, with four MPI processes. 

By assigning visibilities to each MPI rank at once, this time ordered with a bucket sort algorithm due to their $v$ coordinate, it is in principle possible to keep data local to each rank and carry out the successive steps of the pipeline without communicating any data. The whole grid is constructed on the basis of the actual $u$ and $v$ coordinates of the data. For simplicity, we kept the one-dimensional slab distribution (extension to two-dimensional distributions will be the subject of a future work).
If a specific rank is assigned a slab between $v_i$ and $v_{i+1}$, it means that data with $v$ coordinate in that interval must be assigned to the rank. Of course data are originally unorderly distributed, so MPI Sendrecv operations are necessary to share out visibilities among the processes.
However, in Figure \ref{fig:original_dd} it is clear that central regions in the $u-v$ plane are much denser of data than the external ones, leading to a strongly unbalanced visibility redistribution, if one naively allocates a uniformly distributed grid. In fact, tasks which are assigned external slabs in the $u-v$ plane would not participate to the gridding to the same extent as tasks which are assigned to central slabs. 

\begin{figure}[h]
    \centering
    \includegraphics[width=0.7\textwidth]{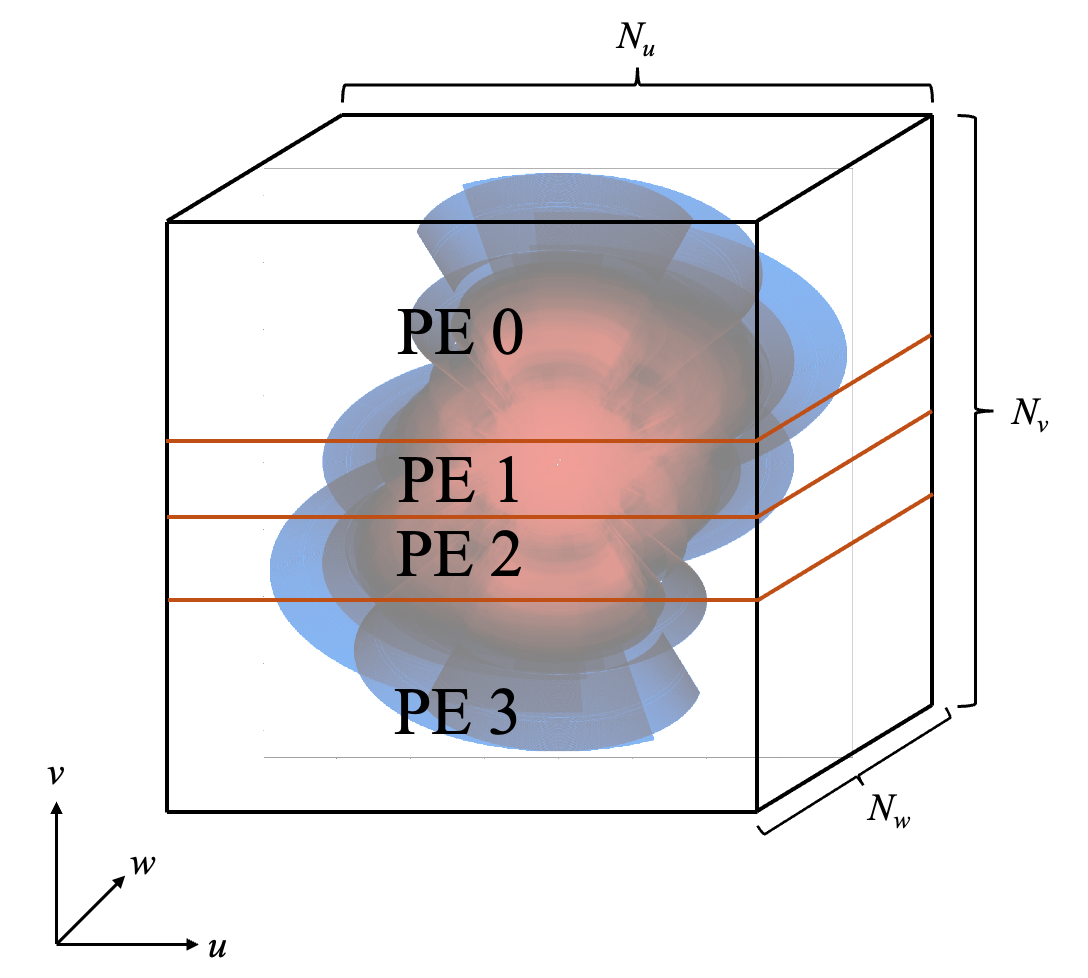}
    \caption{New domain decomposition in rectangular slabs along the $v$ axis, with four MPI processes. This time the domain distribution is non even since due to the higher density of point in the central regions of the $u-v$ plane.}
    \label{fig:new_dd}
\end{figure}

To get a balanced gridding imaging step, a non trivial domain decomposition is compulsory, with finer partitions in central $u-v$ regions and coarser partitions in external ones. Figure \ref{fig:new_dd} is an example of what would happen with four MPI processes. Due to the lower density of points in external regions of $u-v$ plane, processes 0 and 3 would have a larger slab, whereas processes 1 and 2 would be assigned a smaller slab due to the higher density of points in central regions. The size of each region is assigned through a one-dimensional Gaussian distribution:

\begin{equation}
    w_v(i_v) = e^{-\frac{(i_v-\mu_v)^2}{2\sigma_v^2}}
    \label{eq:gaussian-dd}
\end{equation}

representing a 1D Gaussian domain decomposition along the $v$-axis for parallel computing applications using MPI. This approach aims to distribute grid points among processes in a non-uniform manner, assigning fewer points to processes at the edges of the domain and more points to processes located at the center. The $u$-axis, by contrast, is not decomposed and is assigned in its entirely to each process.

The function that performs the decomposition takes as input the total dimensions of the grid ($N_u$, $N_v$), the total number of processes ($P$) and the current rank of the process (rank).
The core of the algorithm involves these steps:
\begin{itemize}
    \item \textbf{$u$-axis allocation:} Each process is assigned the full length of the $u$-axis.
    \item \textbf{Gaussian weight calculation:} For the $v$-axis, the function first calculates the center ($\mu_v$) and standard deviation ($\sigma_v$) of a non-normalized Gaussian distribution. The standard deviation is scaled by a \textit{sigma factor}, hereafter $\sigma$, a parameter that controls the ``spread" of the Gaussian curve. A loop then computes a ``weight" for each point $i_v$ from $0$ to $N_v - 1$ using the Gaussian formula (Equation \ref{eq:gaussian-dd}).
    \item \textbf{Domain boundary determination:} The total weight is then divided into partitions of equal weight $P$. The function iterates through the points $i_v$, accumulating the weights, and sets the boundary points ($v_{boundaries}$) where the cumulative weight for a partition is reached. This method ensures that each process receives a sub-domain with an equal workload in terms of ``weight", even though the number of points in each sub-domain varies.
    \item \textbf{Local sub-domain assignment:} Finally, the $v_i$ and $v_{length}$ for the current process are determined by looking up its assigned boundaries from the $v_{boundaries}$ array based on its rank. This gives the process the starting index and size of its local subdomain along the $v$ -axis.

    Since visibilities must be spread out during gridding on a two-dimensional mesh necessary to apply a Fast Fourier Transform algorithm, one must consider that boundary points might be associated to more than one MPI task. This shrewdness led to the inclusion of \textbf{ghost regions}, which is given by half the convolution kernel dimension, normalized to grid size (i.e. the image pixel size):
    \begin{equation}
        \varepsilon = \frac{(Kernel-1)/2}{N_x}
        \label{eq:kernel}
    \end{equation}
\end{itemize}
This means that if one process is to be assigned data with $v_i < v < v_{i+1}$, the actual data assigned will be $v_i - \varepsilon < v < v_{i+1} + \varepsilon$.

\subsection{Gridding}
\label{gridding-step}
The data redistribution based on the non-uniform domain decomposition described in Section \ref{dd-theory} leads to a much more balanced gridding operation step. We stress out that the \rick~version of Paper I, II and III performed the gridding operation just after I/O ops, and since data were read evenly from the Measurement Set, data distribution was uniform among all MPI tasks. The current version relies on adaptive grid distribution, with central regions thinner than external ones. This decomposition allows the number of visibilities per process to be more balanced. The balance is not perfect, and an approach based on one-dimensional histogram decomposition is being developed to obtain an even better data distribution and make the gridding stage ideally balanced, taking into account the ghost regions defined in Equation \ref{eq:kernel}.

\subsection{Fast Fourier Transform}
\label{heffte-theory}
The non-uniform grid distribution must be supported by a Fast Fourier Transform that allows for such a decomposition. FFTW\footnote{\url{https://www.fftw.org/}} only supports configurations for meshes evenly distributed among all MPI processes involved, and it must be ruled out for this application. The HeFFTe library\footnote{\url{https://icl-utk-edu.github.io/heffte/index.html}} allows for non-uniform one-dimensional, two-dimensional and three-dimensional decompositions. Furthermore, slab, pencils and even bricks decompositions are supported by the library \citep{ayala2020heffte}. 

The library utilizes already existing FFT implementations under the hood, ensuring portability across several HPC platforms, depending on a backend which is defined at compile time. In fact, CPU FFTs are supported through the FFTW backend, i.e. HeFFTe wraps to FFTW library for actual plan creations and calculations. However, the library is fully portable to GPU architectures, with backends wrapping to cuFFT\footnote{\url{https://docs.nvidia.com/cuda/cufft/}}, rocFFT\footnote{\url{https://rocmdocs.amd.com/projects/rocFFT/en/latest/index.html}}, oneMKL FFT\footnote{\url{https://www.intel.com/content/www/us/en/docs/onemkl/developer-reference-c/2023-2/fft-functions.html}}, for NVIDIA, AMD and Intel GPUs, respectively. This solution is extremely powerful for the purposes of this paper, because it allows to define multiple mesh decompositions to catch up with future code developments, for instance for the histogram data redistribution mentioned in Section \ref{gridding-step}, and for the goal of portability, since the code the library runs in many architectures by simply defining the backend at compile time. The code described in Paper II relied on FFTW for CPUs, and on cuFFTMP\footnote{\url{https://docs.nvidia.com/cuda/cufftmp/index.html}} for NVIDIA GPUs, which was not portable. For this study, the GPU-GPU communication needed by FFT is handled by the host and managed internally in the library. This introduces an extra layer of overhead due to internal CPU-GPU memory movements. The choice of the library is also motivated due to the current lack of other implementations supporting distributed (multi-node) GPU FFT.

\section{Methodology and \rick~tests}
\label{sec:methodology}
This section describes the HPC cluster that has been utilized to run the tests, and the description of each run's configuration. We have carried out small tests and large tests. Small tests have been performed with a MeerKAT dataset~\citep[from][]{botteon2025} of $48GB$ memory requirement, whereas large tests have been performed with a LOFAR dataset (from De Rubeis et al., \textit{submitted}) of $363GB$ memory requirement. Due to the fact that domain decomposition is not uniform, i.e. both visibilities and grid are non-evenly distributed, taking the time average among all the MPI tasks is not fair, since the runtime will be dominated by the slowest MPI task. For this purpose, throughout all the paper, the time as measured by the slowest MPI task in each \rick~component will always be taken. So far, \rick~does not perform any cleaning or deconvolution, so all the images that we are going to represent in this paper will always be dirty images.

\subsection{LEONARDO}
\label{th:leonardo}
\subsection*{System Overview}
The \textbf{LEONARDO} supercomputer is a Tier-0, pre-exascale system of the EuroHPC Joint Undertaking (JU), hosted by CINECA at the Bologna Technopole in Italy. It is a BullSequana XH2000 system provided by Atos and designed for both traditional High-Performance Computing (HPC) and Artificial Intelligence (AI) workloads. The system is composed of two main compute partitions: the accelerated \textbf{Booster Module} and the general-purpose \textbf{Data-Centric General Purpose (DCGP) Module}. In this paper tests, we will focus only on the accelerated partition.
\begin{itemize}
\item \textbf{Peak Performance (Total):} $\sim250$ PetaFLOPS (PFLOPs) High-Performance LINPACK (HPL)
\item \textbf{AI Performance:} $\sim10$ ExaFLOPS (EFLOPs) (FP16)
\item \textbf{Total Compute Nodes:} 4992 (3456 Booster + 1536 DCGP)
\end{itemize}
\subsubsection{Booster Module}
The Booster Module is designed for maximum computational acceleration, particularly for AI and heavily parallelized HPC applications.
\begin{itemize}
\item \textbf{System Type:} BullSequana X2135 "Da Vinci" single-node GPU Blade
\item \textbf{Number of Nodes:} 3456
\item \textbf{CPU per Node:} 11× Intel Xeon Platinum 8358 (Ice Lake), 32 cores at 2.6 GHz
\item \textbf{GPU per Node:} 13× NVIDIA custom Ampere A100 GPU, 64 GB HBM2e14
\item \textbf{Inter-GPU Link:} NVIDIA NVLink 3.0, $200 GB/s$ bidirectional bandwidth per pair
\item \textbf{System Memory (RAM):} 512 GB DDR4 3200 MHz16
\item \textbf{Network Interconnect:} 17× dual-port NVIDIA Mellanox HDR100 ConnectX-6 NICs ($\sim400 Gbps$ aggregated bandwidth per node)
\item \textbf{HPL Performance (Module):} $\sim240 PFLOPs$
\end{itemize}
\subsubsection{Interconnect}
The partitions are interconnected via a high-speed network designed for low latency and high data throughput.
\begin{itemize}\item \textbf{Technology:} NVIDIA Mellanox HDR InfiniBand ($200 Gb/s$)
\item \textbf{Topology:} Hierarchical cell-based architecture with a \textbf{Dragonfly+} topology, utilizing NVIDIA Quantum QM8700 Smart Switches.
\item \textbf{Connectivity:} The system is organized into $\sim23$ cells, with a non-blocking two-layer fat-tree topology within each cell and an all-to-all connection between the cells (inter-group interconnection).
\end{itemize}
\subsubsection{Compiler}
For all the tests performed in this paper, the code has been compiled with the NVC-24.3 compiler, where, in the cases of GPU offloading, OpenMP GPGPU has been used. 

\subsection{Small tests}
\label{small_tests}
\subsubsection{MeerKAT data}
\label{sec:meerkat}
The different configurations utilized to carry out the small tests are listed in \ref{table:compute_units}. They refer to the reference computational unit (\CU) of the configuration. In this case we will always refer to an entire node, running the pure MPI CPU-only tests, the hybrid MPI+OpenMP CPU-only tests, and eventually the MPI+OpenMP CPU+GPU tests. The image grid size has been set up to $8192^2$ pixels, with a $32$ $w$ planes, for a grid memory requirements of $32GB$. We could have gone further with grid memory size, but we had to catch up with MeerKAT interferometric resolution, meaning that finer pixel partitions would have led to unphysical sources representations. Increasing the $\sigma$, introduced in Equation \ref{eq:gaussian-dd}, parameter minimizes grid decomposition imbalance across MPI ranks, thereby mitigating overhead during FFT and w-correction stages. This represents a critical trade-off: lower $\sigma$ values result in an FFT-bound execution, whereas higher values shift the bottleneck toward gridding operations. Empirical benchmarks indicate that $\sigma = 1000$ serves as the optimal threshold for current requirements. Nevertheless, more sophisticated decomposition algorithms are under development to dynamically determine the ideal $\sigma$ based on visibility distribution and available MPI resources.

\begin{table}[t]
\begin{center}
\Large
\centering \tabcolsep 3pt
\resizebox{0.70\columnwidth}{!}{%
\begin{tabular}{l|c|c|c|c|c|c|c|c}
\hline
& Configuration & MPI tasks & OpenMP threads & GPUs & $\sigma$ \\ 
\hline
CPU & MPI & 32 & 1 & 0 & 1000\\
\hline
CPU & MPI + OpenMP & 4 & 8 & 0 & 1000\\
\hline
CPU + GPU & MPI + OpenMP (GPU) & 4 & 8 & 4 & 1000\\
\hline
\end{tabular}
}
\end{center}
\caption{Definition of different \CUs~utilized as reference for different configurations strong scaling tests.}
\label{table:compute_units}
\end{table}

\begin{figure}[h]
    \centering
    \includegraphics[width=0.7\textwidth]{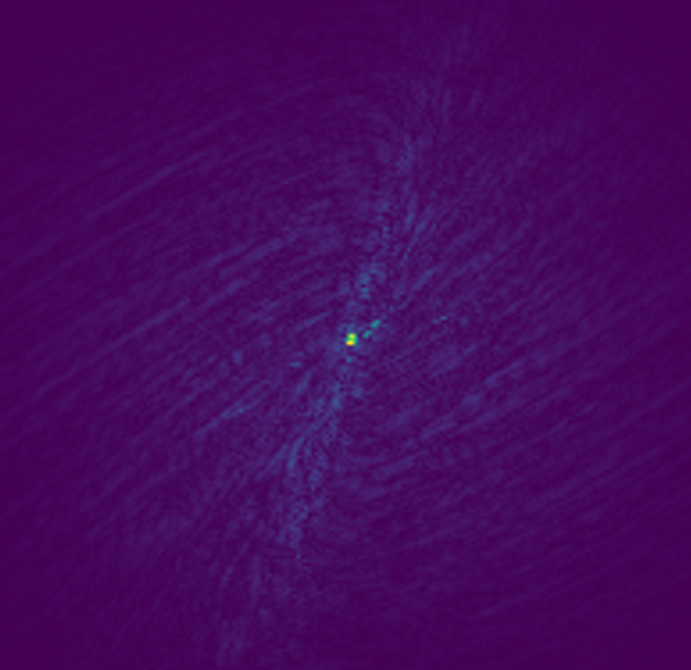}
    \caption{Dirty image produced with \rick~starting from MeerKAT data for the Ophiuchus galaxy cluster~\citep[from][]{botteon2025}}
    \label{fig:meerkat}
\end{figure}

The resulting image is shown in Figure \ref{fig:meerkat}. A central source, associated to the Ophiuchus galaxy cluster~\citep{botteon2025}, is distinguishable in the image centre. This is a strong validation test for \rick~since MeerKAT is a SKA-Mid precursor. The lack of fluxes scale is due to a current deviation between \rick~fluxes and WSClean ones. We are already working to validate also the code's physical scale with the homologous result coming from WSClean, which is taken as the reference one. The difference between RICK and WSClean’s fluxes is due to a rescaling procedure on the u,v,w coordinates needed in RICK, since casacore is still not incorporated into the code. We will get rid of this normalization difference as soon as casacore will be included in the pipeline and the code will be able to read MS directly. For this reason, the absolute flux scale is currently not consistent with the ones from other state of the art codes, but it will be in the future versions of RICK. As stated before, the current offsets in absolute flux scale are not compatible with realistic values, therefore any statement on the image quality would be not valid. To compare the results with the older \rick~version, we will always go for natural weighting. 

\subsubsection{Strong scaling tests}
\label{sec:meerkat_strong}
\begin{figure}[h!]
    \centering
    \includegraphics[width=0.7\textwidth]{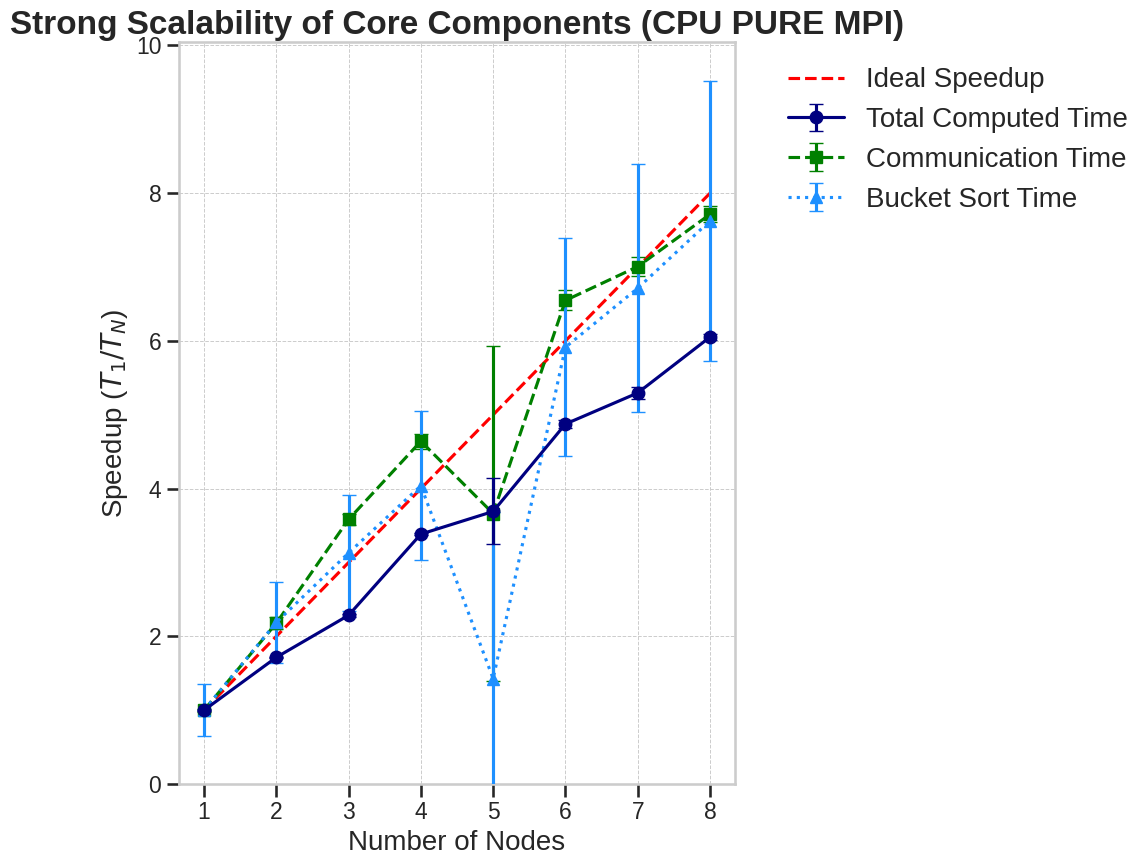}
    \includegraphics[width=0.7\textwidth]{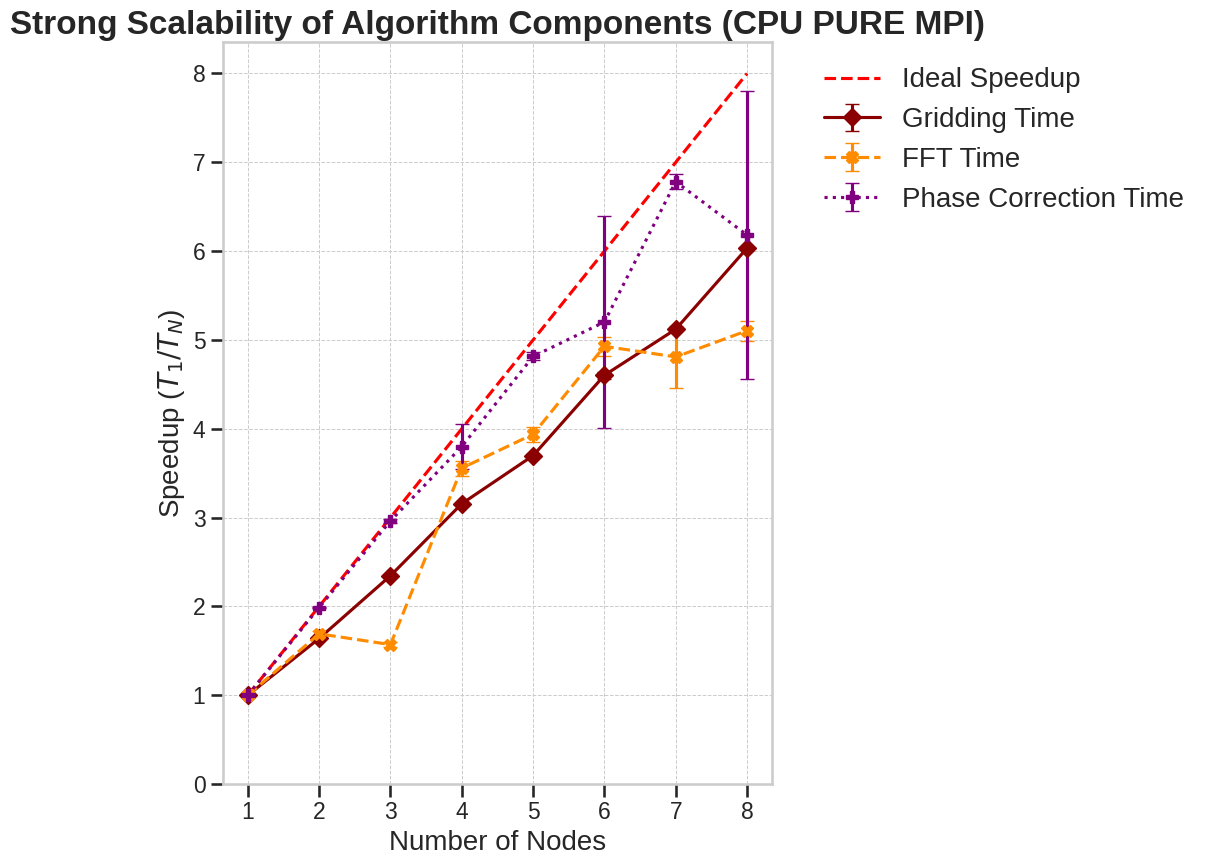}
    \caption{Strong scaling of the different \rick~components for the pure MPI configurations. Upper panel: strong scalability of core components. Total computed time is the whole core runtime from which I/O has been subtracted, communication time is the time for visibility redistribution and bucket sort time is the time which takes the visibility sorting algorithm along the $v$-axis. Lower panel: strong scalability of algorithmic components. Times for gridding, FFT and phase correction algorithmic steps.}
    \label{fig:pure_mpi_speedup}
\end{figure}

By keeping the visibilities and grid size fixed, strong scaling tests have been carried out for the three configurations described above, ranging from 1 \CU\ up to 8 \CUs. The speedup, defined as $\frac{T_1}{T_N}$, i.e. the time needed for 1 \CU~runs, divided by the time needed for $N$ \CUs~runs. Every run has been performed 5 times, and then averages and errors using the standard deviations have been plotted. Results for the pure MPI configurations are shown in Figure \ref{fig:pure_mpi_speedup}. To have a better visualization, the different code components have been split into two plots. The upper panel shows the total computed time, the communication time, the bucket sort speedups, respectively. What is called in the plots the ``Total Computed Time" is the code runtime in which the I/O has been ruled out, since it will be the subject of a future paper. I/O performance is currently constrained by two main factors: \textit{i)} hardware limitations in the CPU-to-filesystem interconnect, where the bottleneck exists at the processor level rather than the individual core level, thereby inhibiting multi-core scaling; and \textit{ii)} internal overheads within the MPI-I/O library, where standard blocking calls trigger synchronizations that impede ideal scaling. Bucket sort is a bit noisy, meaning that there is a non-trivial standard deviation among the different runs, while communication and runtime are pretty stable. The important achievement is the communication speedup, which is a linear speedup (even superlinear) with few \CUs. This is one of the two the main goals we have achieved, since in Paper II and Paper III the communication impact became worse and worse as the number of \CUs~increased. This is due to the point-to-point coupled communications, in which smaller and smaller buffers are allocated as the number of MPI tasks increases, resulting in much quicker data distribution. In the previous case, independently of the number of MPI tasks, the entire grid was to be communicated unavoidably, eventually leading to the observed bottleneck and the $96\%$ communication impact in runtime observed in Paper III. The lower panel plots the speedup of the most computing intensive parts of the code, gridding, FFT and phase correction, respectively. The three components are balanced in this specific configuration, as noticed by the good scaling observed. The FFT worse scaling is explained with two considerations: it is not embarrassingly parallel, due to the inner communications unavoidably needed by the algorithm, and at one point the $8192^2$ grid is not large enough to guarantee a good scalability with up to $256$ MPI tasks, as in the largest runs with 8 \CUs. 

\begin{figure}[h!]
    \centering
    \includegraphics[width=0.7\textwidth]{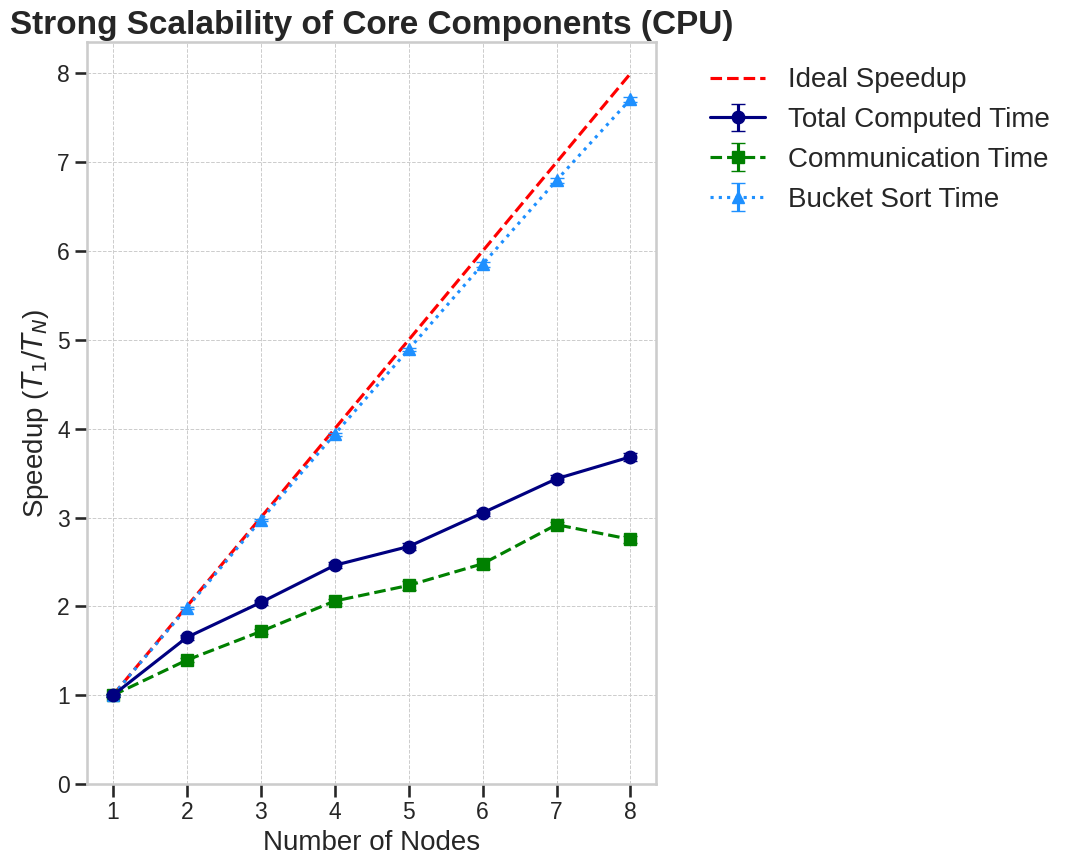}
    \includegraphics[width=0.7\textwidth]{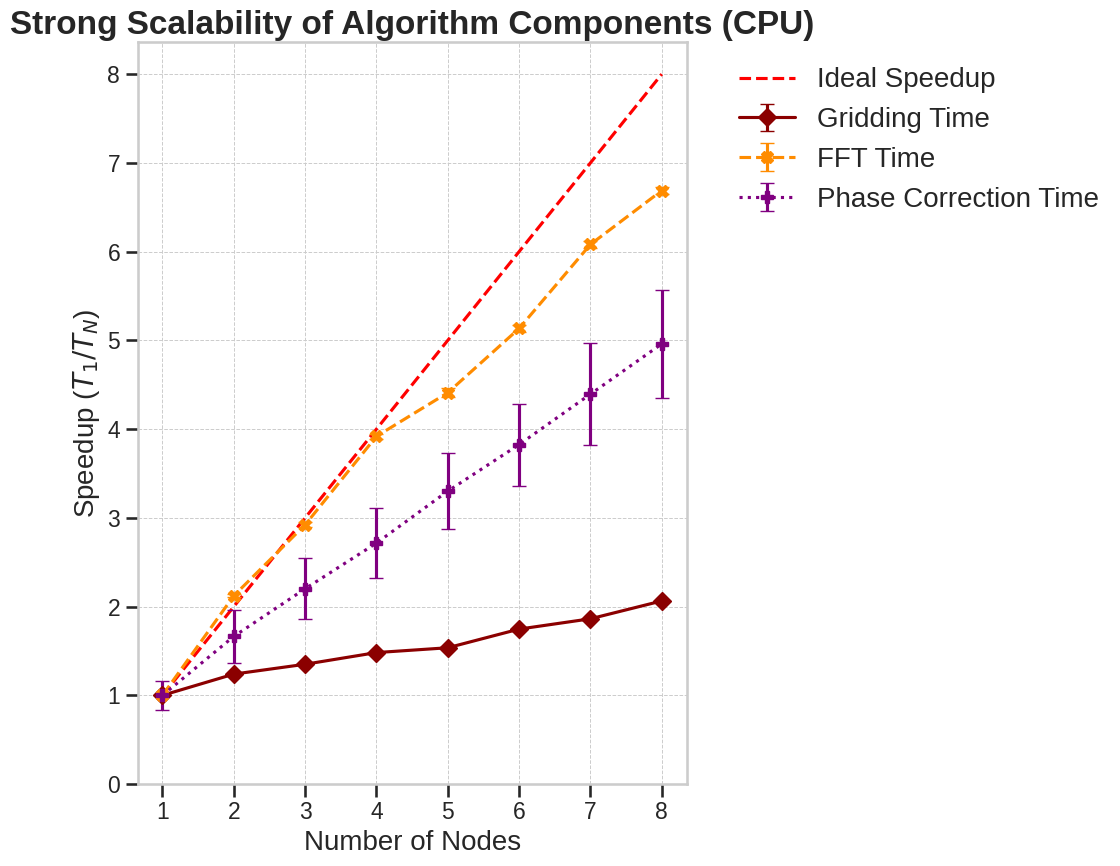}
    \caption{Same as Figure \ref{fig:pure_mpi_speedup}, but for MPI+OpenMP configurations.}
    \label{fig:omp_speedup}
\end{figure}

Speedups for strong-scaling tests for MPI+OpenMP CPU-only configurations are plotted in Figure \ref{fig:omp_speedup}. The philosophy is the same as the one in Figure \ref{fig:pure_mpi_speedup}. Looking at the upper panel, this time the bucket sort is less noisy and almost perfectly scaling, while the computing runtime and the communication display now a worse scaling. This is due to the much less number of MPI tasks, which are now 4 per nodes, instead of 32 as in the previous case. The communication scalability signals that this time visibilities are more unbalanced, meaning that data rearrangement requires larger communication buffers with central region tasks performing much more MPI communications than external region tasks. This visibility unbalance reflects also in the lower panel, especially looking at gridding scaling. It is easy to see that with a few MPI tasks per node, visibilities turn out to be strongly unbalanced, with processes pertaining to the central regions carrying out more communications and computations. FFT scales better this time, since a strongly unbalanced gridding unavoidably leads to a better balanced grid, i.e FFT. Phase correction term scales worse than FFT. Different compilers can exhibit different behaviour, due to the inherently distinct OpenMP support.

\begin{figure}[h!]
    \centering
    \includegraphics[width=0.7\textwidth]{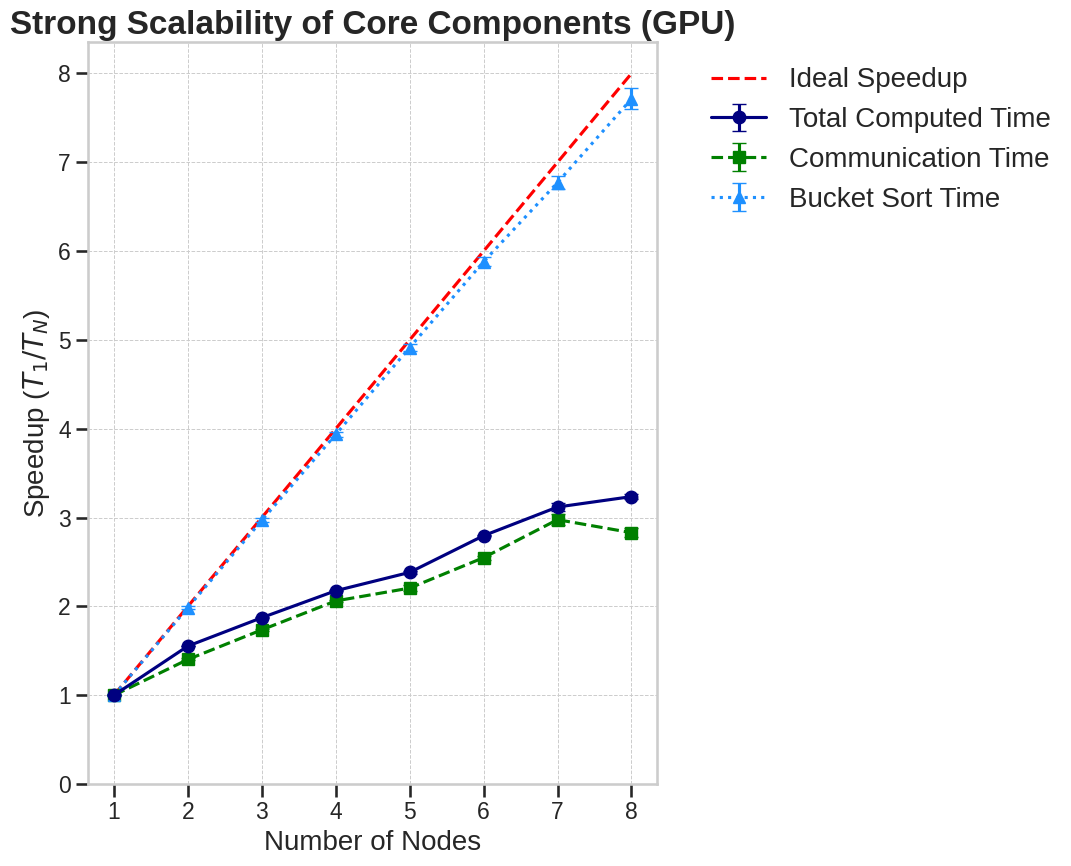}
    \includegraphics[width=0.7\textwidth]{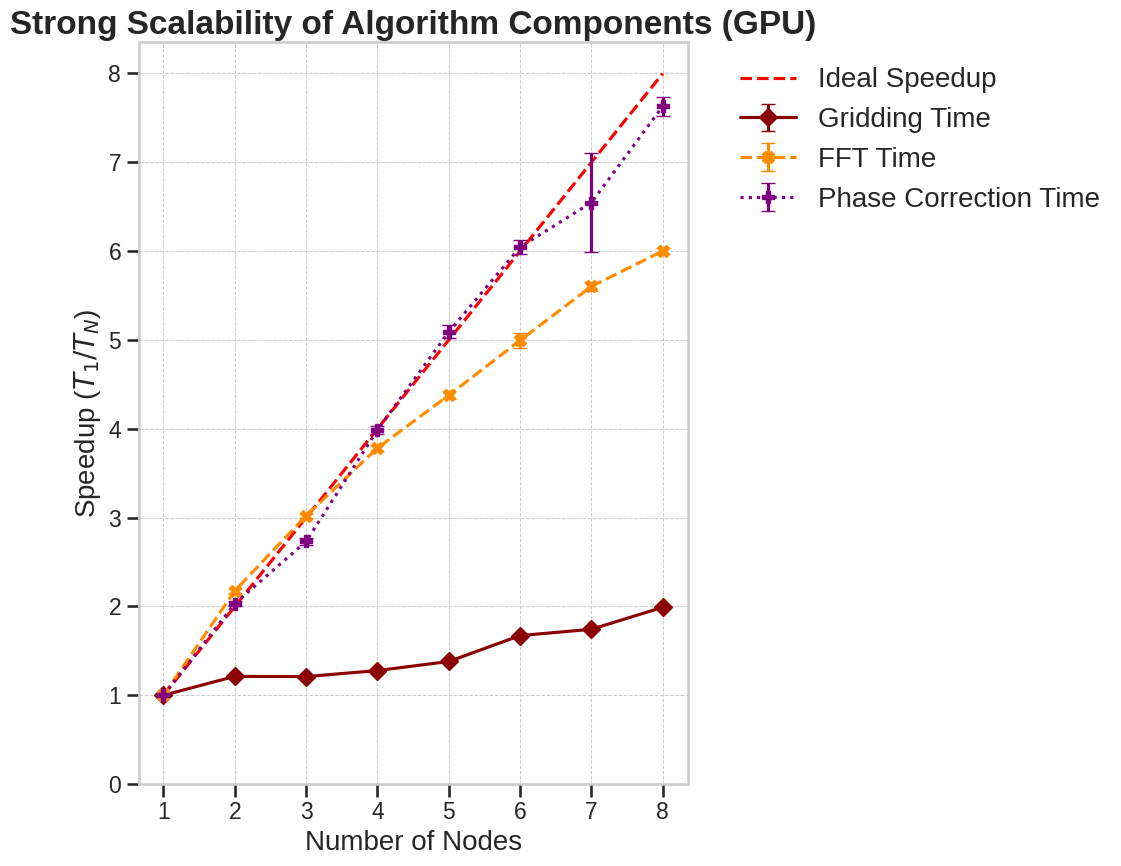}
    \caption{Same as Figure \ref{fig:pure_mpi_speedup}, but for GPU configurations.}
    \label{fig:gpu_speedup}
\end{figure}

Results for GPU runs are shown in Figure \ref{fig:gpu_speedup}. In the upper panel communication and bucket sort speedups are the same as in Figure \ref{fig:omp_speedup}, this is because none of the two algorithms has been offloaded yet. However, computing time scales differently since the offloaded components, shown in lower panel, give now a different contribution. In fact, even though gridding scaling is behaving similarly to the OpenMP CPU case (the number of MPI tasks is the same and visibilities are as well unbalancedly distributed), both FFT and phase correction now show significantly different speedups. FFT scales linearly with a few \CUs, while its efficiency starts decreasing from 5 \CUs~on. Although host-mediated communication introduces non-negligible CPU-GPU latency at scale, the high computational throughput of the GPUs currently offsets this overhead. As the CU count increases, the reduced data volume per MPI rank helps prevent the interconnect from becoming a critical bottleneck. To address the scaling limits of host-mediated transfers, integration of GPU-Direct (RDMA) is underway, which will enable peer-to-peer GPU communication and significantly minimize latency in future iterations. Phase correction scalability shows that the algorithm is best suited for GPU offloading and that the NVIDIA compiler has a better OpenMP support for GPUs than the one for multi-core CPUs. 

Real timings in seconds (averaged over 5 runs) and corresponding errors for 1 \CU, 2 \CUs~and 8 \CUs~are listed in Table \ref{table:best_timings_combined}. Showing the tables for all \CUs~would have been redundant. It is clear that GPUs are much faster than the other configurations. In the Table, gridding is $\sim3.4$ times faster than the MPI+OpenMP configuration, and a dramatic $\sim10$ factor faster than the pure MPI case. The huge difference between the two CPU-only configurations is explained as follows. They're both using the same number of cores, but in a completely different way. In the MPI case, one has 32 independent processes with a hugely unbalanced visibility distribution. External regions tasks have negligible points, whereas the central region tasks have to do all the work. With 4 MPI tasks per node, the visibility distribution is not so unbalanced, so there are not MPI tasks which are not working, but all of them have a comparable workload. Also, each task spawns 8 OpenMP threads performing the gridding step in a second level of parallelism, resulting in a dramatic time decrease compared to the pure MPI runs. Of course, with many \CUs, the huge number of MPI tasks takes over, even if originally the visibility distribution was unbalanced, and at one point the gridding with MPI becomes better than the MPI+OpenMP one. With 2 \CUs, GPUs are faster by factors of $\sim7.3$ and $\sim3.3$ over MPI and CPU-only MPI+OpenMP runs, respectively. With 8 \CUs, GPUs are faster by factors of $\sim3.2$ and $\sim3.3$ over the two CPU-only configurations. The visibility unbalance is huge in 1 \CU~for MPI runs, but becomes less pronounced for many \CUs, whereas exactly the opposite turns out to happen for MPI+OpenMP (CPU and GPU) configurations, as the grid becomes more balanced at the expense of visibilities. GPUs are $\sim1.6$, $\sim2.1$ and $\sim1.95$ times faster in FFT than pure MPI, and $\sim4.6$, $\sim4.6$ and $\sim4.1$ times faster than MPI+OpenMP case, for 1, 2 and 8 \CUs, respectively. Eventually, GPUs show a runtime gain in phase correction by factors of $\sim4.6$ and $\sim18.5$ on 1 \CU, by factors of $\sim4.6$ and $\sim22.2$ on 2 \CUs~and by factors of $\sim6.3$ and $\sim32.3$ on 8 \CUs. This confirms that phase correction is a better algorithm to exploit GPUs, but also that NVIDIA compiler has a better OpenMP support for offloading than multi-core, as mentioned above.               

\begin{table}[t]
\begin{center}
\centering \tabcolsep 3pt
\resizebox{\textwidth}{!}{%
\Large
\begin{tabular}{l|c|c|c|c|c|c}
\hline
\multicolumn{7}{c}{\textbf{1 \CU}} \\
\hline
\textbf{Configuration} & \textbf{Runtime} & \textbf{Communication} & \textbf{Bucket sort} & \textbf{Gridding} & \textbf{FFT} & \textbf{Phase corr.} \\
\hline
MPI & $73.77 \pm 0.29$ & $10.24 \pm 0.14$ & $0.66 \pm 0.16$ & $49.69 \pm 0.04$ & $11.96 \pm 0.12$ & $1.20 \pm 0.01$\\
\hline
OpenMP & $68.25 \pm 0.75$ & $10.99 \pm 0.11$ & $2.12 \pm 0.01$ & $17.21 \pm 0.02$ & $33.10 \pm 0.21$ & $4.82 \pm 0.56$\\
\hline
GPU & $25.64 \pm 0.02$ & $10.99 \pm 0.01$ & $2.12 \pm 0.01$ & $5.02 \pm 0.01$ & $7.24 \pm 0.03$ & $0.26 \pm 0.01$\\
\hline
\multicolumn{7}{c}{\textbf{2 \CUs}} \\
\hline
\textbf{Configuration} & \textbf{Runtime} & \textbf{Communication} & \textbf{Bucket sort} & \textbf{Gridding} & \textbf{FFT} & \textbf{Phase corr.} \\
\hline
MPI & $42.99 \pm 0.20$ & $4.69 \pm 0.15$ & $0.30 \pm 0.01$ & $30.31 \pm 0.06$ & $7.07 \pm 0.09$ & $0.60 \pm 0.05$\\
\hline
OpenMP & $41.33 \pm 0.42$ & $7.88 \pm 0.07$ & $1.07 \pm 0.01$ & $13.88 \pm 0.15$ & $15.60 \pm 0.06$ & $2.89 \pm 0.40$\\
\hline
GPU & $16.5 \pm 0.02$ & $7.83 \pm 0.04$ & $1.07 \pm 0.01$ & $4.14 \pm 0.01$ & $3.33 \pm 0.03$ & $0.13 \pm 0.01$\\
\hline
\multicolumn{7}{c}{\textbf{8 \CUs}} \\
\hline
\textbf{Configuration} & \textbf{Runtime} & \textbf{Communication} & \textbf{Bucket sort} & \textbf{Gridding} & \textbf{FFT} & \textbf{Phase corr.} \\
\hline
MPI & $12.19 \pm 0.07$ & $1.33 \pm 0.01$ & $0.08 \pm 0.00$ & $8.24 \pm 0.01$ & $2.34 \pm 0.04$ & $0.19 \pm 0.01$\\
\hline
OpenMP & $18.53 \pm 0.03$ & $3.99 \pm 0.04$ & $0.27 \pm 0.01$ & $8.34 \pm 0.01$ & $4.95 \pm 0.01$ & $0.97 \pm 0.03$\\
\hline
GPU & $7.92 \pm 0.07$ & $3.88 \pm 0.06$ & $0.27 \pm 0.01$ & $2.52 \pm 0.01$ & $1.20 \pm 0.01$ & $0.03 \pm 0.00$\\
\hline
\end{tabular}
}
\end{center}
\caption{Real timings (in seconds) averaged over 5 runs with relative errors for each configuration on 1 \CU, 2 \CUs~and 8 \CUs.}
\label{table:best_timings_combined}
\end{table}

\subsection{Large tests}
\label{large_tests}

\subsubsection{LOFAR data}
\label{sec:lofar}
\begin{figure}[h]
    \centering
    \includegraphics[width=0.7\textwidth]{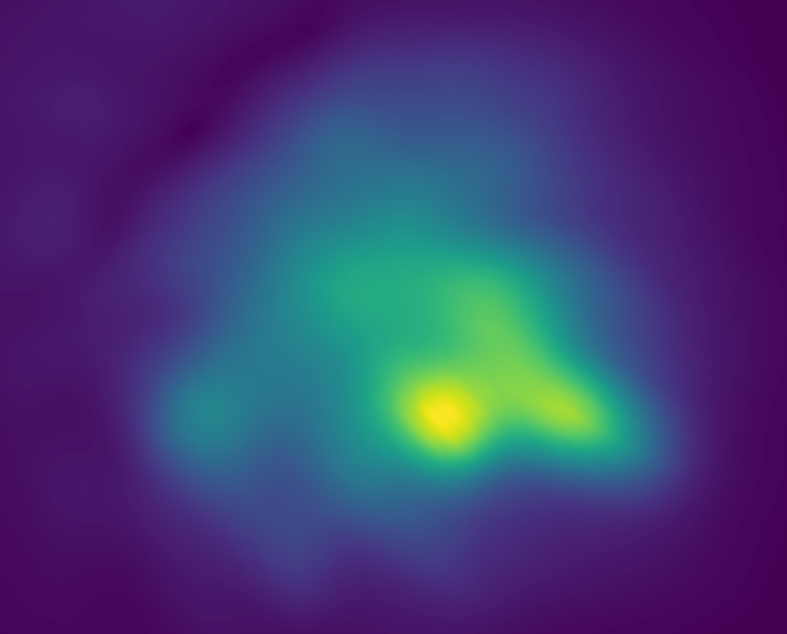}
    \caption{Dirty image produced with \rick~ starting from radio galaxies observed with LOFAR.}
    \label{fig:lofar}
\end{figure}

Aside from small tests with MeerKAT data, we stressed out the communication and FFT thanks to much larger LOFAR data ($363GB$ of visibilities). The image is shown in Figure \ref{fig:lofar}. LOFAR is a SKA-Low precursor, and this result encourages our work since the code has been validated by producing images with real data from both SKA-Mid and SKA-Low precursors. Again, the fluxes scale is not present for the dirty image (see Section \ref{sec:meerkat}).

\subsubsection{Strong scaling tests}
\label{sec:lofar_strong}
The large test setup consists in one-shot runs with a MPI+OpenMP pure CPU configuration and a GPU configuration, each one with 256 \CUs. The image pixel resolution is $65536^2$, with $128$ $w$-planes. This choice stresses out the FFT, due to the large pixel resolution and the number of planes, since one FFT per plane is to be carried out. The single \CU~configurations are the same as the ones in Table \ref{table:compute_units}, in the second and third lines, respectively. Multiple shots have not been performed due to the important run durations which leads to extreme core hours consumption. Results are listed in Table \ref{table:best_timings_large}. Overall, using GPUs for such large tests leads to a $\sim2$ speedup factor, communication and bucket sort are supposed to be the same since those code portions have not been ported to GPUs, even if there is a slight fluctuation in communication time between the two runs (17 seconds over 10 minutes), but what cares the most are the $\sim11.2$ speedup factor in gridding, the $\sim19$ speedup factor in FFT and the dramatic $\sim42.8$ speedup factor in phase correction, when GPUs are used. We stress out that there is no domain decomposition difference between the two configurations, since the same number of MPI tasks have been spawned in each run. Again, gridding and phase correction behave very well on GPUs, with this huge difference compared to multi-core OpenMP again partially due to a better GPU than multi-core offloading for the NVIDIA compiler. Here we discuss the second main achievement of this paper. Since FFT is an algorithm which is commonly used in astrophysics and cosmology and not only in radio imaging, such an advantage when using the HeFFTe (which is also portable across several architectures) leads an important achievement which can be tested in many different applications. Furthermore, we expect that one of the future HeFFTe releases will implement the GPU-GPU Remote Memory Access high-speed connection, having a large impact in the communication which currently passes through the host. The 10 minutes communication time, in which visibilities are redistributed among all the tasks, means that we did not manage to remove it, however, compared to the large runs presented in Paper II, we experience more than one order of magnitude improvement (on the CPUs), with the communication passing to a couple of hours to essentially minutes, with the chance of scaling when more and more processes are used, which was the limit in the original code implementation. 

\begin{table}[t]
\begin{center}
\centering \tabcolsep 3pt
\resizebox{\textwidth}{!}{%
\Large
\begin{tabular}{l|c|c|c|c|c|c}
\hline
\multicolumn{7}{c}{\textbf{256 \CUs}} \\
\hline
\textbf{Configuration} & \textbf{Runtime} & \textbf{Communication} & \textbf{Bucket sort} & \textbf{Gridding} & \textbf{FFT} & \textbf{Phase corr.} \\
\hline
OpenMP & $1057.69$ & $599.95$ & $0.08$ & $212.93$ & $549.42$ & $6.43$\\
\hline
GPU & $511.57$ & $617.84$ & $0.08$ & $18.99$ & $28.90$ & $0.15$\\
\hline
\end{tabular}
}
\end{center}
\caption{Real timings (in seconds) for single shot runs in OpenMP and GPU configurations on 256 \CUs.}
\label{table:best_timings_large}
\end{table}

\section{Conclusions}
\label{sec:conclusions}
In this paper, we have presented \rick~2.0 (i.e. the new version of \rick), an accelerated and portable pipeline for radio interferometric imaging, designed to meet the extreme computational demands of SKA precursors and the future SKA. We have successfully validated the pipeline's correctness and performance using real observational data from MeerKAT (an SKA-Mid precursor) and LOFAR (an SKA-Low precursor).

The most significant advancement in this work is the integration of the HeFFTe library, which provides two fundamental advantages. First, it ensures portability, allowing \rick~to be deployed efficiently across heterogeneous systems, from multi-core CPUs to GPUs. Second, it delivers exceptional performance, especially on GPUs, proving to be highly effective for computationally intensive scenarios involving large pixel resolutions and a high number of frequency planes.

Furthermore, as hinted at in the discussion of large-scale tests with LOFAR, the FFT is a fundamental and ubiquitous computational operation in many fields of astrophysics and cosmology. The adoption of a portable and high-performance library like HeFFTe, especially for GPUs, thus represents a strategic advantage that extends beyond the specific application of radio imaging. This approach paves the way for future optimizations in a wide range of scientific codes that rely on distributed Fourier transforms.

This new implementation directly addresses and solves the critical bottlenecks identified in our previous analyses (Paper II, Paper III). In that work, we demonstrated that communication overheads became the dominant performance limiter, for large images and large computing resources, consuming up to $96\%$ of the total runtime. This limitation was inherent in the algorithm, which required the communication of the entire data grid and thus could not scale. \rick's new architecture fundamentally resolves this issue, dramatically reducing the communication impact and enabling true scalability for next-generation radio astronomy challenges.

With the primary computational and communication bottlenecks now addressed, I/O operations may emerge as the next performance-limiting factor. Future work will focus on optimizing data handling. A promising avenue for further enhancement is the implementation of asynchronous techniques to overlap I/O operations with communication, effectively hiding latency and further improving the overall time-to-solution. One possibility is to implement the ADIOS2 library \citep{GODOY2020100561}, which supports asynchronous I/O and does not show performance drops for non-trivial domain decompositions. Much effort is being put on validating \rick's fluxes against other imaging tools (i.e. WSClean). The results of this work will be presented in a scientific paper. \rick~provides a robust, scalable, and accelerator-ready solution, representing a critical step forward in preparing imaging software for the data deluge of the SKA, with the purpose of large-scale operational use in radio astronomy.

\section*{Acknowledgment}
Supported by Italian Research Center on High Performance Computing Big Data and Quantum Computing (ICSC), project funded by European Union - NextGenerationEU - and National Recovery and Resilience Plan (NRRP) - Mission 4 Component 2 within the activities of Spoke 2 (Fundamental Research and Space Economy), (CN 00000013 - CUP C53C22000350006). EDR is supported by the Fondazione ICSC, Spoke 3 Astrophysics and Cosmos Observations. National Recovery and Resilience Plan (Piano Nazionale di Ripresa e Resilienza, PNRR) Project ID CN\_00000013 \enquote{Italian Research Center for High-Performance Computing, Big Data and Quantum Computing} funded by MUR Missione 4 Componente 2 Investimento 1.4: Potenziamento strutture di ricerca e creazione di \enquote{campioni nazionali di R\&S (M4C2-19)} - Next Generation EU (NGEU).
This work is also supported by  "Bando per il finanziamento della Ricerca Fondamentale 2024 dell’Istituto Nazionale di Astrofisica" Decreto n. 8/2024 MiniGrant "Numerical Chaos in Cosmological Simulations"
We acknowledge the CINECA, for the availability of high performance computing resources and support. The HPC tests and benchmarks this work is based on, have been produced on the Leonardo Supercomputer at CINECA (Bologna, Italy) in the project: INA24\_C3T03.
\bibliographystyle{elsarticle-harv} 
\bibliography{biblio}

\end{document}